\documentclass[%
 reprint, 
superscriptaddress, 
 amsmath, amssymb, 
 aps, 
]{revtex4-2}
\usepackage[]{pifont}%星号
\usepackage{svg} % 核心包，依赖graphicx
\usepackage{graphicx}% Include figure files
\usepackage{dcolumn}% Align table columns on decimal point
\usepackage{bm}% bold math
\usepackage{xcolor} % for color text
\usepackage{comment}
\begin{document}

\preprint{APS/123-QED}

\title{Robust Unidirectional Edge States in the Continuum in non-Topological Floquet Photonic Crystals}

\author{Hai-rong Huo}
\affiliation{School of Physics, Beijing Institute of Technology, Beijing 100081, 
China}
\author{Bingsuo Zou}
\affiliation{MOE \& Guangxi Key Laboratory of Processing for Non-Ferrous Metals and Featured Materials, School of Physical Science and Technology, Guangxi University, Nanning 530004, China}
\author{Yongyou Zhang}
\email[Corresponding author: ]{yyzhang@bit.edu.cn}
\affiliation{School of Physics, Beijing Institute of Technology, Beijing 100081, 
China}


\date{\today}% It is always \today, today, 

\begin{abstract}
Robust unidirectional edge propagation is conventionally attributed to topological protection. Whether edge states in the continuum (EICs) can exhibit such robustness in non-topological systems remains an open question. Here we demonstrate robust unidirectional EICs in Floquet photonic crystals (PhCs) composed of a honeycomb lattice of helical waveguides, where both time-reversal and spatial inversion symmetries are broken. Within a topologically trivial parameter regime of this system, where the Chern, valley Chern, and winding numbers all vanish, the EIC robustness is decoupled from topology. Instead, the robustness originates from a $z$-periodic Floquet artificial gauge field geometrically locked to the helical lattice. Numerical simulations show that the EIC survives 120$^\circ$-bent edges, ${\sim}6\%$ on-site potential noise, and ${\sim}27\%$ hopping phase noise. This work establishes a paradigm for robust light propagation in non-topological systems and broadens the physical basis for unidirectional EICs.
\end{abstract}

\maketitle

\section{Introduction}

Topological band theory establishes that robust unidirectional edge states arise from topological invariants, such as chiral electronic edge states in the quantum Hall effect \cite{PhysRevLett.45.494, alma991013977539706535}. Systems hosting these topologically protected edge states are fundamentally immune to backscattering from non-magnetic impurities and structural imperfections, rooted in their topological invariants \cite{PhysRevLett.49.405}. This landmark discovery underpins the field of topological insulators \cite{PhysRevLett.95.146802, doi:10.1126/science.1148047, PhysRevLett.98.106803, PhysRevB.75.121306, PhysRevB.79.195322, hsieh_topological_2008} and provides a foundational platform for quantum technologies \cite{PhysRevX.4.041022, doi:10.1073/pnas.1810003115, liu_spin-filtered_2014, mellnik_spin-transfer_2014, fan_electric-field_2016}.
Photonic topological insulators subsequently emerged as a direct extension of topological band theory from condensed-matter physics \cite{haldane_possible_2008, wang_observation_2009}. Leveraging the mature theoretical framework and photonic fabrication techniques, this concept has been realized in diverse optical platforms \cite{lu_2014_nature_topological, Tang2022LASER_review_TopologicalPhotonicCrystals}, including gyromagnetic systems \cite{wang_observation_2009}, coupled resonators \cite{hafezi_robust_2011, fang_realizing_2012}, bianisotropic metamaterials \cite{khanikaev_photonic_2013}, and waveguides \cite{rechtsman_photonic_2013}.

Bound states in the continuum (BICs) --- non-radiative localized states embedded within the radiation continuum --- represent a rapidly advancing research area enabling precise light control. While the pioneering proposal by von Neumann and Wigner \cite{von_neumann_uber_1993} remains experimentally unrealized, numerous distinct BIC types have been demonstrated in a broad range of material systems \cite{PhysRevLett.102.167404, Valle:08, McIver_1996, suh_displacement-sensitive_2003, longhi_floquet_2013, craster_dynamic_2013,hsu_2016_NatReview_bound,Azzam_2021_AOM_photonic}. PhCs are a particularly attractive platform for exploring and engineering BICs due to their mature structural design capabilities \cite{PhysRevLett.109.067401, suh_displacement-sensitive_2003, hsu_observation_2013, Kang2021PRLMerging, Sang2024NANOPHOTONICSAchievingquasi-bound, Wu2024NATURECOMMUNICATIONS_Exciton, Schiattarella2024NATUREDirective, Wu2024PhysRevB.109.085436Momentum}. In particular, topologically protected BICs \cite{Zhen_2014_PRL_Topological, Hu2021LIGHT_SCIENCEHigher_order, De2023AOMHalf-IntegerTopological} exhibit robustness to perturbations in system parameters, endowing them with significant application in photonic cavities \cite{Barczyk_2022_LASER_Interplay, Chen_2022_SCIENCEBULLETIN_Observation}, polarization conversion \cite{Guo_2017_PRL_PolarizationConversion, Guo_2019_Adv_Arbitrary, Kang_2022_Nat.Comm._Coherent}, and quantum control \cite{poddubny_2018_arxiv_nonlinear,Zhang_2025_COMMUTP_Phase}.

\begin{figure}[t]
\includegraphics[width=0.48\textwidth]{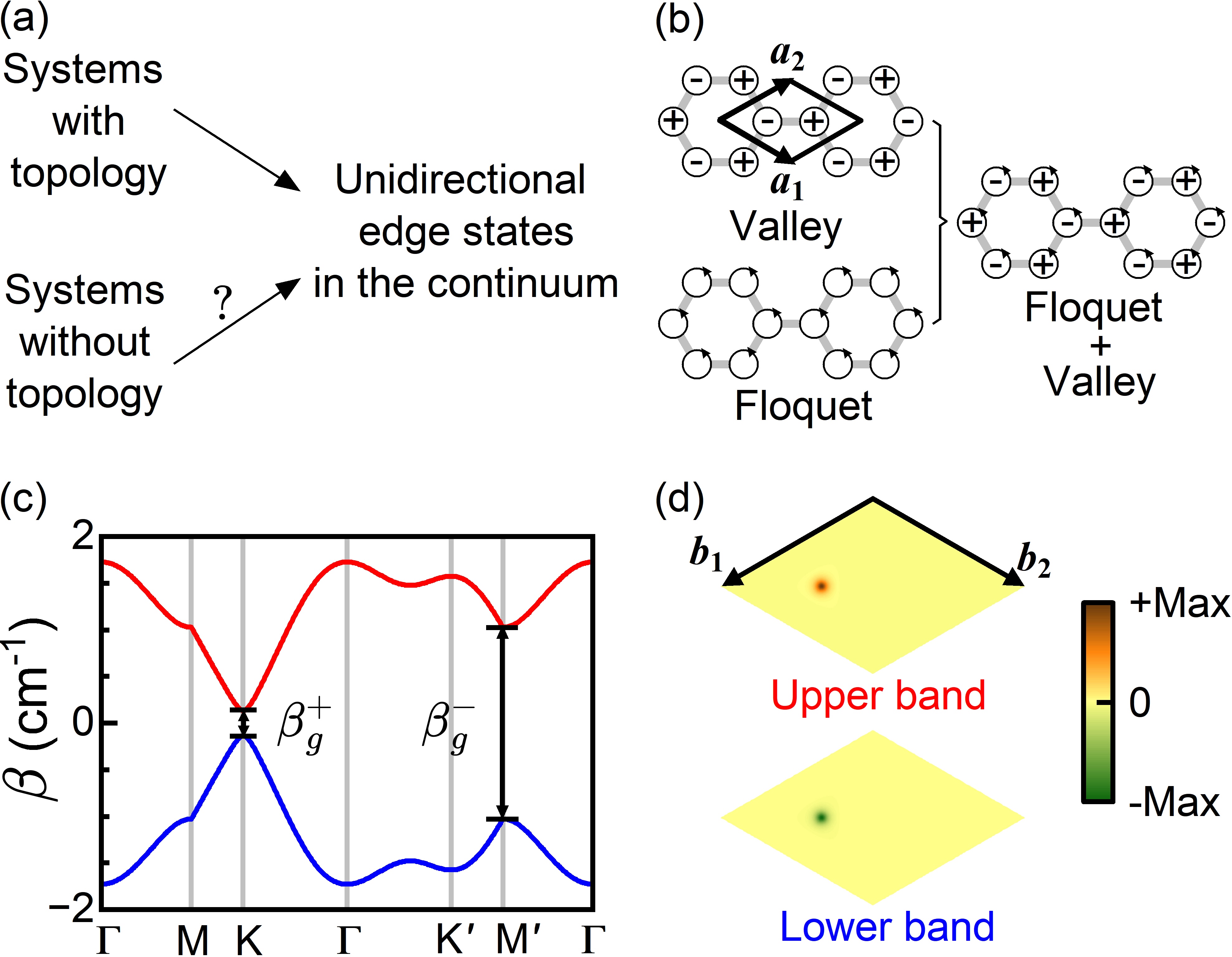}
\caption{(a) Origin of robust unidirectional EICs. 
(b) Schematic of valley (sublattice potential) and Floquet (helical waveguide) degrees of freedom in a honeycomb lattice and their combination. $\bm{a}_1$ and $\bm{a}_2$ are the lattice vectors with length $a$; $\pm$ denotes the sublattice on-site potentials of $\pm\delta$. Helical waveguide orientation is indicated by the
arrows, out of screen. 
(c) Quasi-energy band structure of the Floquet PhC for $R=6\,\mu{\rm m}$ and $\delta=1\,{\rm cm}^{-1}$. (d) Berry flux of the quasi-energy bands in (c), where $\bm{b}_1$ and $\bm{b}_2$ are the reciprocal lattice vectors. Throughout this work, other fixed parameters are $a = 32.9$ $\mu$m, $c=1.9\,{\rm cm}^{-1}$, $n_0=1.45$, $\lambda=0.633\,\mu{\rm m}$, and $\Omega=3.8\pi\,{\rm cm}^{-1}$ \cite{Bandres_2016_PRX_Topological,stutzer_2018_Nature_photonic, Chen_2023_PRB_Kitaev}.}\label{fig1}
\end{figure}

As the edge-state analogue of BICs, edge states in the continuum (EICs) have received comparatively limited attention, despite their untapped potential in applications such as quantum control and fundamental studies such as BIC physics \cite{Zhang2021PRABoundTopologicalEdgeState, feng_2022_OL_boundvalley, sato_2024_NC_observation}.
In topological photonics, bound topological EICs \cite{Zhang2021PRABoundTopologicalEdgeState} and bound valley EICs \cite{feng_2022_OL_boundvalley} have been proposed by leveraging the BIC physics. These topologically protected EICs exhibit unidirectional propagation and intrinsic disorder immunity [see Fig.~\ref{fig1}(a)]. Recently, bound EICs have also been experimentally observed in non-topological photonic crystals (PhCs) consisting of truncated silicon pillars \cite{sato_2024_NC_observation}. However, the lack of topological protection makes such non-topological EICs highly susceptible to disorder-induced perturbations during propagation. The fragility of the non-topological EICs motivates the question of whether it is possible to realize robust EICs with unidirectional propagation and inherent disorder resilience in purely non-topological systems, which is schematically illustrated in Fig.~\ref{fig1}(a).

We show here that such robust unidirectional EICs can indeed be established in Floquet PhCs. Throughout this work, `continuum' refers to the bulk states rather than the radiative extended states above the light cone \cite{Cerjan_2020_PRL_Observation, Benalcazar_2020_PRB_bond, Yin_2024_SB_Valley, Jalali_2026_PRR_Dispersive}. Robustness of these EICs is decoupled from topological protection, since our design simultaneously breaks both time-reversal and spatial inversion symmetries, and the parameter regime supporting these edge states exhibits vanishing Chern, valley Chern, and winding numbers. Perturbation simulations on the Floquet PhCs verify that the mode resilience arises from a $z$-periodic Floquet artificial gauge field. Physically, the unidirectional EICs derive their robustness from a Floquet artificial gauge field that arises from the helicity of the waveguides. That is, this gauge field is geometrically locked to the helical trajectory and is not readily destroyed by moderate perturbations.

\section{Model}

\subsection{Tight-binding bands}

The Floquet PhC under study consists of a honeycomb lattice of helical waveguides, as indicated by out-of-screen arrows in Fig.~\ref{fig1}(b). Experimentally, this structure can be fabricated via local refractive index modulation by precisely controlling laser writing power \cite{rechtsman_photonic_2013}. Evanescent coupling between adjacent waveguides is described by the tight-binding model \cite{rechtsman_photonic_2013},
\begin{equation}
i\partial_{z}\psi_{n}(z) = \delta_n \psi_{n}(z) + \sum_{\langle m\rangle}c\, e^{i\bm{A}(z)\cdot\bm{r}_{mn}}\psi_{m}(z), 
\label{eq:tb}   
\end{equation}
with details provided in Sec.~SI of the Supplemental Material (SM). Here, $\psi_n$ denotes the field amplitude in the $n$-th waveguide, $c$ is the nearest-neighbor evanescent coupling strength, and $\bm{r}_{mn}$ is the displacement vector from the $m$-th to $n$-th waveguide. The on-site potential $\delta_n$ assumes values of $+\delta$ and $-\delta$ for the two inequivalent honeycomb sublattices [marked $\pm$ in Fig.~\ref{fig1}(b)]. The Floquet artificial gauge potential $\bm{A}(z) = A_0[\sin(\Omega z), -\cos(\Omega z), 0]$, encoding the periodic transverse displacement of the helical waveguides, is introduced via Peierls substitution, with $\Omega$ the helix frequency along the optical propagation axis $z$. Here, the amplitude $A_0 = 2\pi n_0 R \Omega/\lambda$ is determined by helix radius $R$, background refractive index $n_0$, and incident wavelength $\lambda$. With $c$ as the reference energy, the band properties of the Floquet PhC are fully determined by $A_0$ and $\delta$. Throughout this work, we fix $a = 32.9$ $\mu$m, $c=1.9\,{\rm cm}^{-1}$, $n_0=1.45$, $\lambda=0.633\,\mu{\rm m}$, and $\Omega=3.8\pi\,{\rm cm}^{-1}$ \cite{Bandres_2016_PRX_Topological, stutzer_2018_Nature_photonic, Chen_2023_PRB_Kitaev}, such that $A_0$ is tuned via $R$.

Nonzero $\delta$ and $R$ explicitly break spatial inversion symmetry and time-reversal symmetry, respectively. Either symmetry enforces the quasi-energy $\beta$ to satisfy $\beta(-\bm{k}) = \beta(\bm{k})$, where $\bm{k}$ denotes the wave vector. By contrast, for the Berry flux $F$, time-reversal symmetry yields $F(-\bm{k}) = -F(\bm{k})$, whereas spatial inversion symmetry gives $F(-\bm{k}) = F(\bm{k})$ (see Sec.~SII and Fig.~S1 of the SM).
However, breaking both symmetries simultaneously invalidates these parity relations, as illustrated in Figs.~\ref{fig1}(c, d). The band structure and Berry flux exhibit distinct asymmetry between the left and right halves of the first Brillouin zone (BZ). This symmetry breaking renders the energy landscapes along high-symmetry paths inequivalent, and the global band minimum, which typically resides at the K point in conventional honeycomb lattices, may shift to the M or \(\Gamma\) point. The Berry flux is freed from any even- or odd-parity constraint in momentum space, see Fig.~\ref{fig1}(d).

\subsection{Phase diagram}

\begin{figure}[t]
\hspace*{-0.01\textwidth}
\includegraphics[width=0.48\textwidth]{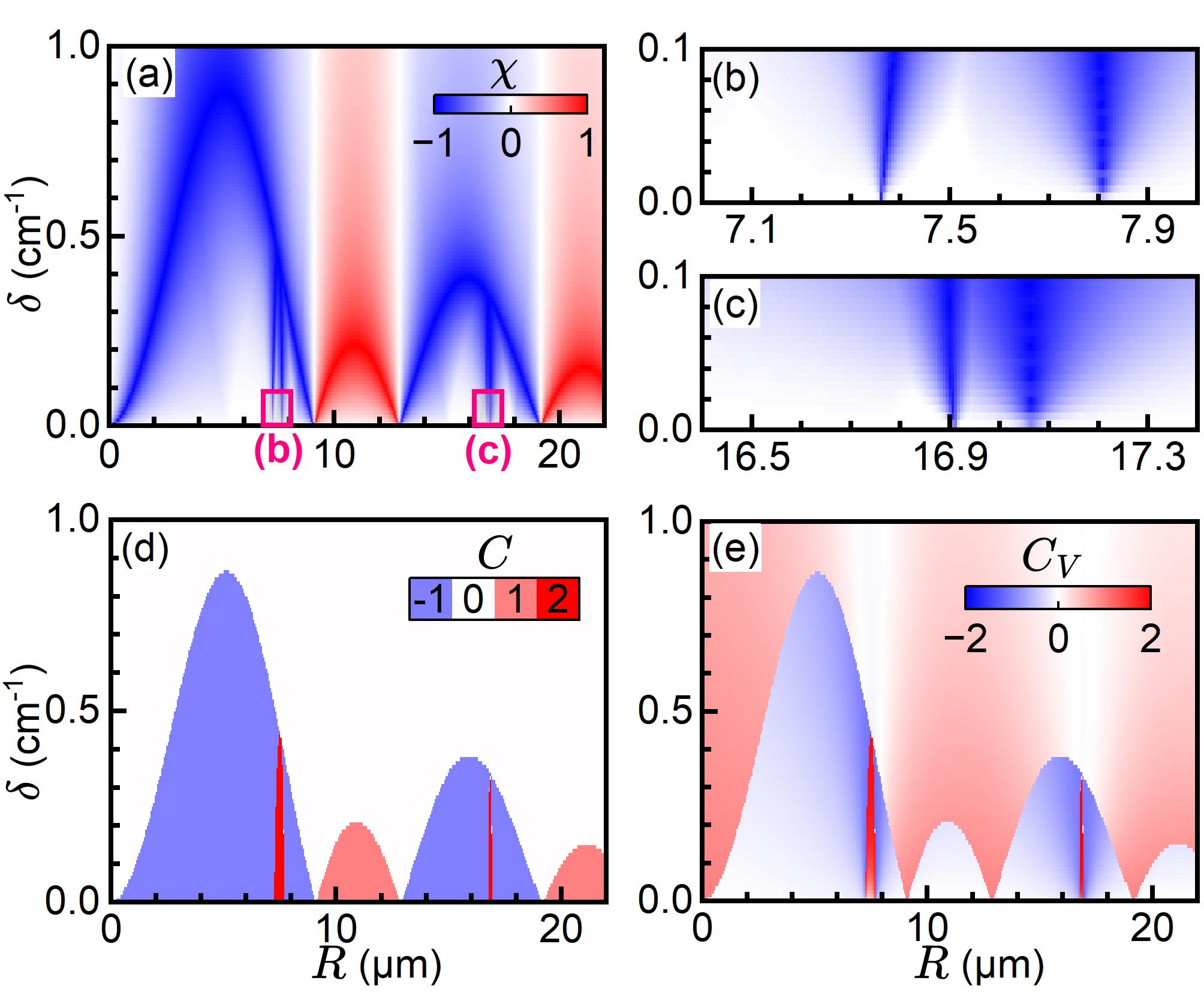}
\caption{(a) Relative quasi-energy band gap $\chi $ as a function of helix radius $R$ and on-site potential $\delta$. (b, c) Enlarged regions denoted in (a) with red boxes. (d) Topological phase diagram in the \(R\)-\(\delta\) space, distinguished by Chern number $C$. (e) Variation of valley Chern number $C_V$ in the \(R\)-\(\delta\) space.\label{fig2} }
\end{figure}

To quantify this momentum-space asymmetry, we introduce the relative band gap \(\chi\), defined as
\begin{equation}
\chi \equiv \frac{\beta_g^+ - \beta_g^-}{\beta_g^+ + \beta_g^-}, 
\label{eq:relative_gap}
\end{equation}
where \(\beta_g^+\) and \(\beta_g^-\) denote the band gaps along high-symmetry paths in the positive and negative halves of the first BZ, respectively. Values of \(\chi = \pm 1\) indicate gap closure for \(\beta_g^-\) and \(\beta_g^+\), respectively. Figure~\ref{fig2}(a), with its red-boxed regions enlarged in Figs.~\ref{fig2}(b) and \ref{fig2}(c), presents \(\chi\) as a function of the helix radius \(R\) and on-site potential \(\delta\). The gap-closing lines in the \(R\)-\(\delta\) space demarcate distinct topological phases, confirmed by the distributions of Chern number \(C\) in Fig.~\ref{fig2}(d) and winding number $W_0$ in Fig.~S2(c) of the SM. The calculation methods are provided in Sec.~SII of the SM. In the studied range of $[0,\ 1.0\ {\rm cm}^{-1}]$, large \(\delta\) values suppress the formation of Chern topological phases.

The valley Chern number \(C_V\) quantifies the valley-Hall topology and is formally defined as the difference of Berry flux integrals over two momentum-space subregions,
\begin{equation}
C_V = \frac{1}{2\pi} \left( \sum_{{\bm k}\in S_+} F(\bm{k}) \, - \sum_{{\bm k}\in S_-} F(\bm{k}) \right), 
\label{eq:valley_chern}
\end{equation}
where \(S_+\) and \(S_-\) typically denote momentum-space regions around the K and K$'$ valleys, respectively. However, in our system, time-reversal symmetry breaking renders the Berry flux non-odd and non-vanishing far from the K and K$'$ points; thus, we redefine \(S_+\) and \(S_-\) as the positive and negative halves of the first BZ for practical calculation. The distribution of \(C_V\) in the \(R\)-\(\delta\) parameter space is plotted in Fig.~\ref{fig2}(e), which exhibits identical boundary contours to the \(C\) distribution in Fig.~\ref{fig2}(d). Next, we will demonstrate in what follows that the unidirectional EICs reported herein originate from neither Chern topology nor valley topology, but their robustness is tied to the helicity of the waveguides, i.e., Floquet artificial gauge field.

\section{Edge states in the continuum}
\subsection{Band structures}

\begin{figure}[t]
\includegraphics[width=0.48\textwidth]{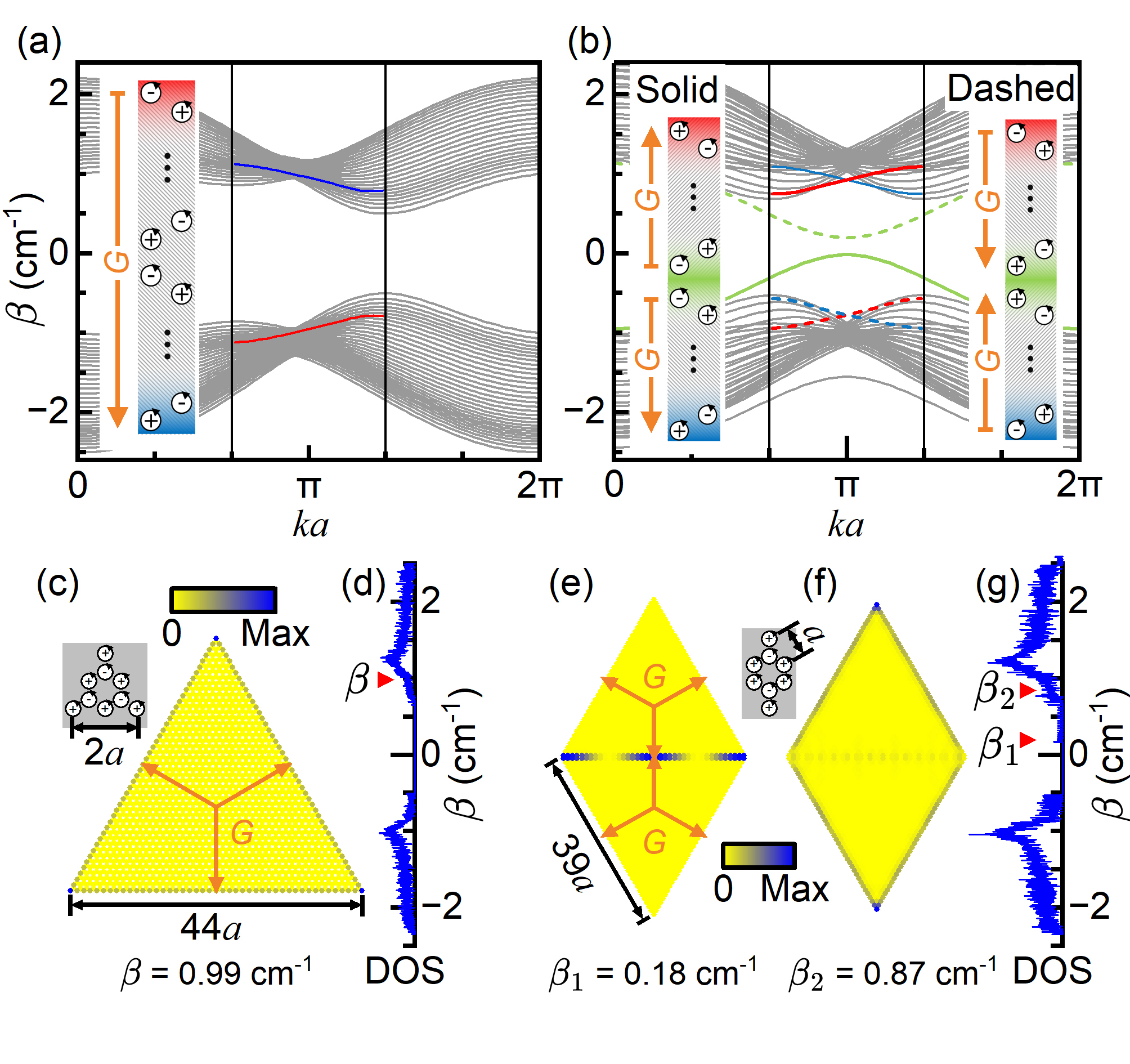}
\caption{Dispersion characteristics and spatial distributions of edge states in Floquet PhCs. (a, b) Band dispersions for Floquet PhCs without and with a valley interface, respectively, with corresponding structural insets and $\bm G$ directions. In (b), solid and dashed curves correspond to the left and right structural insets, respectively. The colored regions in the insets of (a) and (b) mark the spatial locations of the corresponding EIC modes. (c) Spatial distribution of the EIC at $\beta=0.99\,\text{cm}^{-1}$ in a equilateral triangular PhC and (d) the corresponding density of states (DOS). (e, f) Spatial distributions of two modes at $\beta_1=0.18\,\text{cm}^{-1}$ and $\beta_2=0.87\,\text{cm}^{-1}$ in a rhombic PhC, and (g) the corresponding DOS profile. The gray shaded triangular and rhombic insets illustrate the lattice geometries adopted in (c-g). The lattice contains 64 sites in (a, b), 2025 sites in (c, d), and 3200 sites in (e-g). All panels adopt identical structural parameters: $R=11.64\,\mu\text{m}$, $\delta=0.8\,\text{cm}^{-1}$, and $G = 4\,\text{cm}^{-2}$.}\label{fig3}
\end{figure}

\begin{figure*}[t]
\centering 
\includegraphics[width=0.98\textwidth]{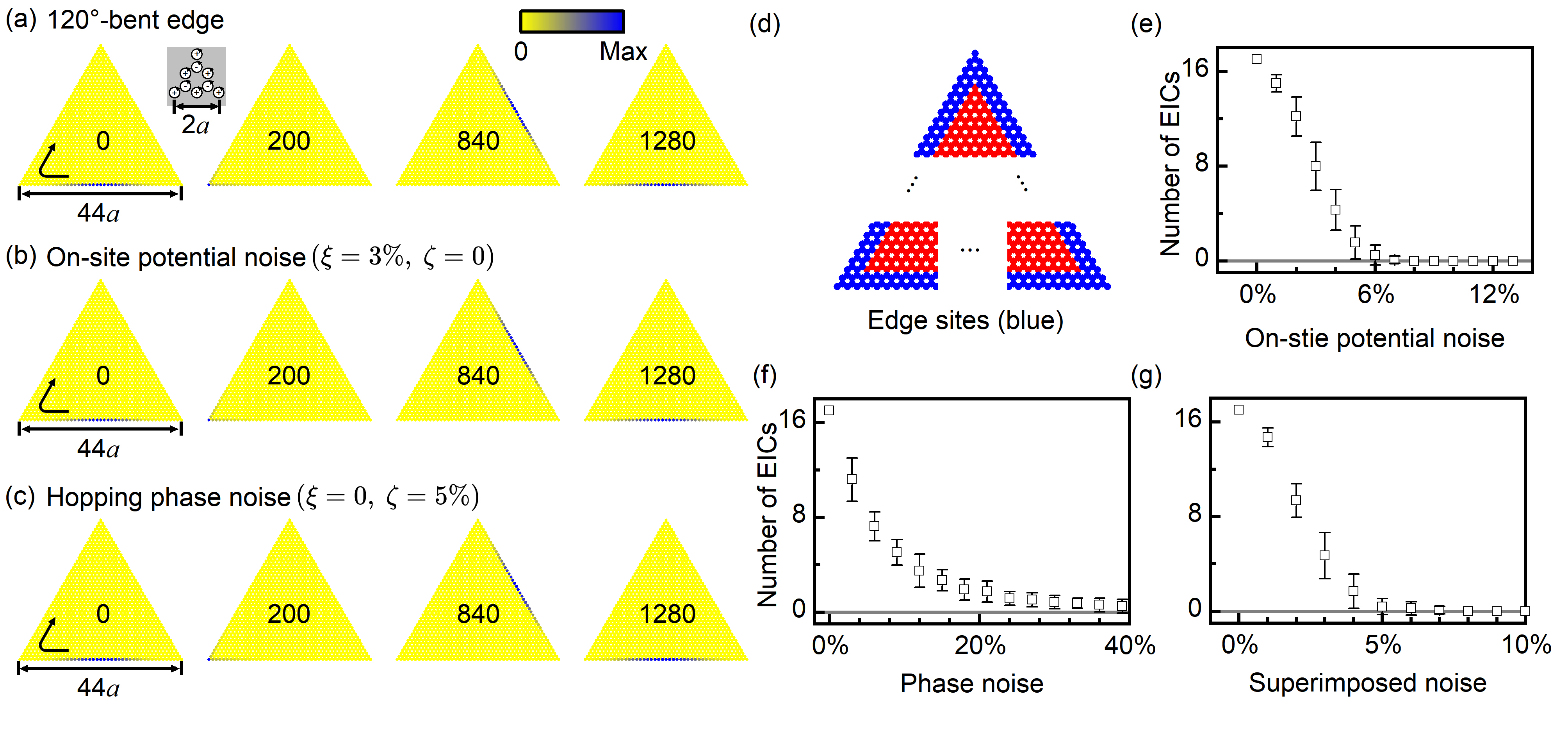}
\caption{Propagation of Gaussian-enveloped EICs in equilateral triangular Floquet PhCs under different perturbation conditions. (a) Bent-corner propagation without disorder ($\xi=\zeta=0$). (b) Bent-corner with on-site potential noise ($\xi=3\%$, $\zeta=0$). (c) Bent-corner with hopping phase noise ($\xi=0$, $\zeta=5\%$). The triangular lattices adopted in (a)–(c) are identical to that in Fig.~\ref{fig3}(c). The EIC at $\beta=0.99\,\text{cm}^{-1}$ is modulated into a Gaussian pulse for all simulations. The numbers 0, 200, 840, and 1280 marked at the lattice centers denote the propagation lengths along the $z$-direction in units of $2\pi\Omega^{-1}$. (d) Schematic of edge sites defined as four layers of blue-colored sites. (e-g) Statistical numbers of sustained EICs as functions of individual on-site potential noise, individual hopping phase noise, and simultaneous dual disorder, respectively.}  \label{fig4}
\end{figure*}

For $R=0$ and $\delta=0$, the Floquet PhC reduces to the tight-binding graphene model, which supports two degenerate zero-energy flat bands localized at zigzag edges \cite{Nakada_1996_PRB_Edge}. Finite sublattice detuning ($\delta\neq0$) opens a bulk gap and splits these flat bands into non-dispersive edge states, and the subsequent introduction of the Floquet artificial gauge field ($R\neq0$) imparts dispersion to these edge states [see Fig.~S8(a) of the SM]. Although these edge states are spectrally isolated from the bulk states in their own momentum regime, they can be driven into the bulk continuum by an extra gradient on-site potential, 
\begin{equation}
\delta_n \rightarrow \delta_n + {\bm r_n} \cdot {\bm G}. 
\label{Gra}
\end{equation}
Here \(\bm G\) denotes the gradient on-site potential direction and \({\bm r_n}\) is the site position. Gradient on-site potential broadens the bulk bands and buries the edge states into the continuum \cite{Li_2022_OL_TopologicalFloquetBIC}, see the red and blue EICs in Fig.~\ref{fig3}(a) for $G=|\bm G| = 4\,\text{cm}^{-2}$. In Fig.~\ref{fig3}(a), the EICs are localized at the top and bottom zigzag edges and exhibit unidirectional propagation, rightward and leftward along the red and blue edges, respectively, referred to the structural inset. The blue EIC has a higher quasi-energy than the red one, because the top-edge sites carry an on-site potential of $-\delta$ while the bottom carry $+\delta$, consistent with $\bm{G}$'s direction. For $R=11.64\,\mu\text{m}$ and $\delta=0.8\,\text{cm}^{-1}$, the Chern number remains \(C=0\), indicating that the gradient potential is too weak to drive a topological phase transition, verified by Fig.~\ref{fig2}(c) and Fig.~S3 of the SM. Accordingly, the observed EICs are not protected by Chern topology. Figure~\ref{fig3}(c), where the equilateral triangular lattice features zigzag-cut boundaries terminated by sites with $+\delta$ (structural inset), presents an illustrative example of such a robust EIC with quasi-energy $\beta=0.99\,\text{cm}^{-1}$. The on-site potential exhibits a gradual increase linearly from the geometric center to the three edges. This EIC resides within the bulk band continuum rather than the band gap, see the density of states (DOS) in Fig.~\ref{fig3}(d).

Since Figs.~\ref{fig3}(a, c, d) do not feature a valley domain wall, we incorporate such a domain wall in Fig.~\ref{fig3}(b) to clarify the correlation between EICs and valley edge states. The valley domain wall is composed of lattice sites with identical on-site potentials, as marked by the green shaded interfaces in the two structural insets of Fig.~\ref{fig3}(b).
The solid colored dispersion curves in Fig.~\ref{fig3}(b) correspond to the left structural inset, whereas the dashed colored curves correspond to the right one. The two structural insets possess opposite on-site potentials, thereby reversing the orientation of the corresponding $\bm G$ vectors.
The red and blue curves represent EICs, while the green curves represent valley edge states. Their spatial distributions are illustrated in the structural insets following the same color coding.
Figure~\ref{fig3}(e) exemplifies a valley edge state in a finite rhombic lattice, which is localized along the valley domain wall terminated by $+\delta$ sites (schematic inset). This valley edge state, with quasi-energy $\beta_1=0.18\,\text{cm}^{-1}$, resides within the photonic band gap [see the DOS in Fig.~\ref{fig3}(g)] and thus exhibits standing-wave characteristics.
In contrast, the mode with quasi-energy $\beta_2=0.87\,\text{cm}^{-1}$ originates from hybridization between EICs and valley edge states, with its spatial distribution in Fig.~\ref{fig3}(f). 
This hybrid mode falls inside the bulk band continuum, see the DOS in Fig.~\ref{fig3}(g). Despite the feasible hybridization between EICs and valley edge states, these two types of modes remain intrinsically distinct. EICs emerge in regions without Berry flux sign reversal and cannot exist at armchair edges. By contrast, valley edge states rely on Berry flux sign reversal across the domain wall and are sustained at armchair edges, see Fig.~S4 of the SM.

\subsection{Robustness}

Despite the absence of topological protection, the EIC shown in Fig.~\ref{fig3}(c) supports robust unidirectional propagation and strong disorder resilience, distinctly differing from topologically trivial edge states and conventional valley edge states. 
To quantitatively characterize such robustness, we modulate the EIC in Fig.~\ref{fig3}(c) using a Gaussian envelope ${1\over\sqrt{\pi\sigma^2}}e^{-x^2/2\sigma^2}$ ($\sigma=9a$) to mimic realistic experimental pulse conditions, as presented in the left panel of Fig.~\ref{fig4}(a). 
In this case, only bent edges are introduced without other noise. 
The Gaussian EIC pulse propagates continuously along the zigzag edge, exhibiting no observable backscattering or mode splitting at $120^\circ$ corners, as confirmed by the field intensity distributions recorded at $z/2\pi\Omega^{-1}=200$, 840, and 1280. 
We attribute this unidirectional propagation to the Floquet artificial gauge field based on two evidences. 
First, the Chern, valley Chern, and winding numbers all vanish at the studied parameter point [see Figs.~\ref{fig2}(c), S2, and S3], ruling out any topological origin. 
Second, such robustness completely vanishes when setting $R=0$, i.e., no Floquet artificial gauge field. 
Physically, the helical waveguide geometry induces a $z$-periodic gauge potential, whose rotating profile enforces unidirectional propagation along the edge.

In Fig.~\ref{fig4}(b), on-site potential noise is introduced at every lattice site via
\begin{equation}
\delta_n \rightarrow \delta_n\cdot[1 + {\rm Rand}(\xi)],
\end{equation}
where ${\rm Rand}(\xi)$ represents a uniform random variable defined over the interval $[-\xi/2,\ \xi/2]$, with $\xi=3\%$ adopted in Fig.~\ref{fig4}(b). 
A direct comparison between Fig.~\ref{fig4}(a) and Fig.~\ref{fig4}(b) demonstrates that the EIC maintains its unidirectional propagation and pulse envelope under moderate on-site noise. 
Such robustness originates from the intrinsic chirality of the Floquet gauge field. The geometric configuration of the helical lattice remains unperturbed under moderate on-site potential fluctuations, thereby sustaining the chiral locking of the EIC mode.
Nevertheless, sufficiently strong on-site noise ($\xi\gtrsim6\%$) can overwhelm this, that is, excessive random potential fluctuations can break the geometric chiral locking mechanism and induce hybridization between the EIC and bulk modes. As a result, the localized EICs on the edges demonstrated in Fig.~\ref{fig4}(d) decay rapidly. This noise-induced degradation is quantitatively verified in Fig.~\ref{fig4}(e), where the number of EICs with an occupation probability exceeding $98\%$ at the blue edge sites in Fig.~\ref{fig4}(d) drops to zero for $\xi\gtrsim6\%$.

To understand the gauge field’s role, we examine the EIC against hopping phase noise via the transformation, 
\begin{equation}
e^{i {\bm A}(z)\cdot {\bm r}_{nm}} \rightarrow e^{i {\bm A}(z)\cdot {\bm r}_{nm}\cdot[1 + {\rm Rand}(\zeta)]},
\end{equation}
where ${\rm Rand}(\zeta)$ is a uniform random variable in $[-\zeta/2, \zeta/2]$. Consistent with the on-site noise results in Fig.~\ref{fig4}(b), Fig.~\ref{fig4}(c) with $\zeta=5\%$ clearly shows that the EIC retains stable unidirectional propagation and pulse envelope under moderate hopping phase noise.
This confirms that the Floquet gauge field is resilient to the hopping phase fluctuations. 
As the hopping phase noise increases to $\zeta\sim27\%$, severe phase disorder disrupts the unidirectional propagation and gradually degrades the EIC robustness. 
Figure~\ref{fig4}(f) quantitatively characterizes this degradation, for which the number of EICs with an occupation probability above $98\%$ at the blue edge sites vanishes for $\zeta\gtrsim27\%$. 

To characterize the joint disorder response, we simultaneously introduce both perturbations with identical strengths, i.e., $\xi=\zeta$.
As shown in Fig.~\ref{fig4}(g), the critical disorder strength at which the number of EICs with an occupation probability exceeding $98\%$ at the blue edge sites vanishes is lower than those observed in the single-disorder cases in Figs.~\ref{fig4}(e) and \ref{fig4}(f).
This indicates that the two disorder sources cooperatively disrupt the EIC robustness. 

Moreover, the case with $G = 0$ is also investigated using identical numerical procedures, and the corresponding results (see Figs.~S5 and S6 of the SM) are consistent with the above discussions on Figs.~\ref{fig3} and \ref{fig4}.
Further examinations of backscattering probabilities under individual local defects (see Figs.~S7–S8 of the SM) still verify the robustness of the proposed EICs.
These disorder simulations confirm that the robust unidirectional propagation of the EIC is governed by the Floquet artificial gauge field, rather than by conventional Chern or valley topological mechanisms. 
It is consistent with the vanishing Chern and winding numbers, as well as the absence of valley domain walls in the system.

\section{Conclusion}

In summary, this work demonstrates robust unidirectional EICs achievable in purely non-topological Floquet photonic crystals. Unlike conventional topological EICs, the proposed mode features vanishing Chern, valley Chern, and winding numbers, confirming its independence from topological protection. Its superior disorder immunity and stable unidirectional propagation stem purely from the $z$-periodic Floquet artificial gauge field, which is geometrically locked to the helical lattice and resistant to moderate perturbations. This work establishes a route for realizing robust unidirectional EICs in non-topological photonic systems and has potential applications in optoelectronic devices and quantum techniques.

\bibliography{ref}% Produces the bibliography via BibTeX.

\end{document}

\preprint{APS/123-QED}

%\title{Supplemental Material for ``Robust Unidirectional Edge States in the Continuum without Topological Protection''}
%\title{Supplemental Material for ``Floquet-gauage-field-induced Photonic Edge States in the Continuum without Topological Protection''}
\title{Supplemental Material for ``Robust Unidirectional Edge States in the Continuum in non-Topological Floquet Photonic Crystals''}

\author{Hai-rong Huo}
\affiliation{School of Physics, Beijing Institute of Technology, Beijing 100081, 
China}
\author{Bingsuo Zou}
\affiliation{MOE \& Guangxi Key Laboratory of Processing for Non-Ferrous Metals and Featured Materials, School of Physical Science and Technology, Guangxi University, Nanning 530004, China}
\author{Yongyou Zhang}
\email[Corresponding author: ]{yyzhang@bit.edu.cn}
\affiliation{School of Physics, Beijing Institute of Technology, Beijing 100081, 
China}


\date{\today}% It is always \today, today, 

\maketitle

%\tableofcontents

%提纲：
%第一段：1、Floquet周期理论；2、以及截断的误差分析。
%第二段：1、当只引入谷的体带；只引入Floquet的体能带。
%第三段：1、贝利曲率的计算的计算公式以及；2、区间内特殊点的能带、贝利曲率；3、windingnumber的计算；4，更大空间范围内的winding数以及chern数
%第四段：1、Zigzag边界态的由来；2、Armchair边界、beard边界、新边界？的一维态。
%第五段：1、对于无序的扰动的讨论，边界态的数目，随无序的变化。2，耗散随时间的变化？
%第六段：

\section{Tight-binding model for Floquet photonic crystals}

The theoretical model of the helical waveguide lattice is constructed based on Maxwell's equations,
\begin{equation}
    \begin{aligned}
    \nabla \cdot \boldsymbol{D} &= 0, \\
    \nabla \times \boldsymbol{E} &= -\frac{\partial \boldsymbol{B}}{\partial t}, \\
    \nabla \cdot \boldsymbol{B} &= 0, \\
    \nabla \times \boldsymbol{H} &= \frac{\partial \boldsymbol{D}}{\partial t}, 
    \end{aligned}
    \label{Maxwell}
\end{equation}
in the absence of free electric currents and free charges. The constitutive relations for dielectric waveguides are given by
\begin{equation}
\boldsymbol{D} = \varepsilon_r\varepsilon_0 \boldsymbol{E}, \quad
\boldsymbol{B} = \mu_0 \boldsymbol{H}.
\end{equation}
where $\varepsilon_r$ is the relative permittivity and $\varepsilon_0$ ($\mu_0$) is the vacuum permittivity (permeability).
Substituting $\boldsymbol{E} = \boldsymbol{E}(x,y,z)e^{-i\omega t}$ and $\boldsymbol{H} = \boldsymbol{H}(x,y,z)e^{-i\omega t}$ into Eq.~\eqref{Maxwell} leads to the Helmholtz equation,
\begin{equation}
\nabla^2\boldsymbol{E} + \varepsilon_r \varepsilon_0 \mu_0 \omega^2 \boldsymbol{E} = 0.
\label{Helmholz}
\end{equation}
For the helical waveguide, we define the lattice plane as the $xy$-plane and the helical direction as the $z$-direction. Since the helical modulation of the waveguide varies slowly compared with the field variations in the $xy$-plane, we can construct the solution by separating out a plane-wave form along the $z$-direction:
\begin{equation}
\boldsymbol{E}(x,y,z) = \boldsymbol{\cal E}(x,y,z)e^{ik_0z}, \label{calE}
\end{equation}
where $k_0 = \frac{2\pi}{\lambda}n_0$. Here, $n_0$ and $\lambda$ are the refractive index of the background material and light wavelength. Substituting Eq.~\eqref{calE} into Eq.~\eqref{Helmholz} yields,
\begin{equation}
\begin{aligned}
i \frac{\partial}{\partial z} \boldsymbol{\cal E}(x,y,z) 
&= -\frac{1}{2k_0} \nabla_t^2 \boldsymbol{\cal E}(x,y,z) 
+ \frac{k_0}{2} \boldsymbol{\cal E}(x,y,z) 
- \frac{\varepsilon \mu \omega^2}{2k_0} \boldsymbol{\cal E}(x,y,z)  
-\frac{1}{2k_0}\frac{\partial^2}{\partial z^2} \boldsymbol{\cal E}(x,y,z),
\end{aligned}
\end{equation}
where $\nabla_t^2=\frac{\partial^2}{\partial x^2}+\frac{\partial^2}{\partial y^2}$. Given the weak variation of the helical waveguide along the $z$-direction, higher-order derivative terms can be neglected, leading to a light propagation equation analogous to the Schrödinger equation \cite{rechtsman_photonic_2013},
\begin{equation}
i \frac{\partial}{\partial z} \boldsymbol{\cal E}(x,y,z) = -\frac{1}{2k_0} \nabla_t^2 \boldsymbol{\cal E}(x,y,z) - \frac{k_0 \Delta n(x,y,z)}{n_0} \boldsymbol{\cal E}(x,y,z),
\label{light_propagation}
\end{equation}
where the refractive index variation
\begin{equation}
    \Delta n = \frac{n^2(x,y,z) - n_0^2}{2n_0}.
\end{equation}

To capture the helicity of waveguides, we further perform a coordinate transformation,
\begin{equation}
\begin{cases}
x' = x + R\cos(\Omega z) \\
y' = y + R\sin(\Omega z) \\
z' = z
\end{cases}
\end{equation}
where $R$ is the helical radius of waveguides and $\Omega$ is the helical frequency along the $z$-direction. The transformed propagation equation reads:
\begin{equation}
    \begin{aligned}
    i \frac{\partial}{\partial z'} \boldsymbol{\cal E}'(x',y',z') 
    &= -\frac{1}{2k_0} \left[ \nabla'_t + i \boldsymbol{A}(z') \right]^2 \boldsymbol{\cal E}'(x',y',z') \\
    &\quad - \frac{k_0 R^2 \Omega^2}{2} \boldsymbol{\cal E}'(x',y',z') \\
    &\quad - \frac{k_0 \Delta n(x',y')}{n_0} \boldsymbol{\cal E}'(x',y',z'),
    \end{aligned}
\end{equation}
where $\boldsymbol{A}(z') = k_0 R \Omega \left[ \sin(\Omega z'), -\cos(\Omega z'), 0 \right]$ is the Floquet gauge potential introduced by the helical geometry of waveguides.
Accordingly, the propagation of light in a honeycomb lattice of helical waveguides can be described by the tight-binding equation,
\begin{equation}
    i\partial_{z}\psi_{n}(z) = \delta_n \psi_{n}(z) + \sum_{\langle m\rangle}c\, e^{i\bm{A}(z)\cdot\bm{r}_{mn}}\psi_{m}(z)
    \equiv \sum_m H_{nm}(z)\psi_{m}(z),
    \label{eq:tb}
\end{equation}
where the prime on $x,y,z$ has been omitted for convenience. The resulted Hamiltonian matrix \(H(z)\) holds \(H(z+Z)=H(z)\) with \(Z=2\pi/\Omega\). By Floquet theory, the general solution to Eq. \eqref{eq:tb} takes the form
\begin{equation}
    \vert\Psi_{\beta}(z)\rangle = e^{-i\beta z}|\Phi_{\beta}(z)\rangle,\quad \vert\Phi_{\beta}(z+Z)\rangle=\vert\Phi_{\beta}(z)\rangle,
\label{solu:floquetstate}
\end{equation}
where \(|\Psi_{\beta}(z)\rangle = [\cdots, \psi_{n-1}(z),\, \psi_{n}(z),\, \psi_{n+1}(z), \cdots ]^{\mathsf{T}} \) is the state vector and \(\beta\) denotes the corresponding quasi-energy.

\newpage 

\section{Calculation ways}

\subsection{Band structure}

To determine the quasi-energies \(\beta\), we transform Eqs. \eqref{eq:tb} and \eqref{solu:floquetstate} into the Fourier domain, yielding the Shirley-Floquet form \cite{Shirley_1965_Phys.Rev._Solution},
\begin{equation}
    \mathcal{H}\vert\boldsymbol{\varphi}_{\beta}\rangle = \beta \vert\boldsymbol{\varphi}_{\beta}\rangle, \quad
    \mathcal{H} = \begin{pmatrix}
    \ddots & \vdots & \vdots & \vdots & \rotatebox{315}{{\vdots}} \\
    \cdots & H^{(0)} - \hbar\omega & H^{(-1)} & H^{(-2)} & \cdots\\
    \cdots & H^{(1)} & H^{(0)} & H^{(-1)} & \cdots \\
    \cdots & H^{(2)} & H^{(1)} & H^{(0)} + \hbar\omega  & \cdots \\
    \ \rotatebox{315}{{\vdots}} & \vdots & \vdots & \vdots & \ddots
    \end{pmatrix}, \quad
    \vert \boldsymbol{\varphi}_{\beta} \rangle= \begin{pmatrix}
    \vdots \\
    \left|\Phi_{\beta}^{(-1)}\right\rangle \\
    \left|\Phi_{\beta}^{(0)}\right\rangle \\
    \left|\Phi_{\beta}^{(1)}\right\rangle \\
    \vdots
    \end{pmatrix},
    \label{eq:Floquet H}
\end{equation}
where \(H^{(m)}\) is the \(m\)-th Fourier component of the time-dependent TB Hamiltonian \(H(z)\) in Eq.~\eqref{eq:tb}. Distinct Floquet eigenstates are confined to the Floquet-Brillouin zone of \([-\Omega/2, \Omega/2]\). The states outside this interval merely correspond to the repetition of those in \([-\Omega/2, \Omega/2]\) \cite{rudner2020floquetengineershandbook}. Consequently, the infinite Hamiltonian \(\mathcal{H}\) can be safely truncated for numerical computations.
In this work, we retain Fourier components up to the ninth order. The Floquet wavefunctions thus take the truncated expansion,
\begin{equation}
    \vert\Psi_{\beta}(z)\rangle =e^{-i\beta z} \sum_{m=-9}^{9} 
    e^{-im\Omega z}\left\vert\Phi_{\beta}^{(m)}\right\rangle
    \label{solu:Fourierpsi}
\end{equation}

To compute the quasi-energy band structure of a Floquet photonic crystal (PhC), we first recast the tight-binding Hamiltonian $H$ into momentum space,
\begin{equation}
    H_{\bm{k}}(z) = \begin{pmatrix}
        \delta & c f_{\bm{k}}\\
        c f_{\bm{k}}^* & -\delta
    \end{pmatrix}, \quad {\rm with\ }
    f_{\bm{k}}= e^{i[\bm{k} + \bm{A}(z)]\cdot \bm{d}_1} + e^{i [\bm{k} + \bm{A}(z) ] \cdot \bm{d}_2} +e^{i (\bm{k} + \bm{A}(z) ) \cdot \bm{d}_3},
    \label{Hk2d}
\end{equation}
where $\bm{d}_i$ ($i=1,2,3$) denote the relative displacement vectors defined in Fig.~\ref{fig1}(b). 

Taking the Fourier components of $H_{\bm k}(z)$ as $H_{\bm k}^{(m)}$ and substituting them into Eq.~\eqref{eq:Floquet H}, we can find the Hamiltonian in momentum space as
\begin{equation}
    \mathcal{H}_{\bm k} = \begin{pmatrix}
    \ddots & \vdots & \vdots & \vdots & \rotatebox{315}{{\vdots}} \\
    \cdots & H^{(0)}_{\bm k} - \hbar\omega & H^{(-1)}_{\bm k} & H^{(-2)}_{\bm k} & \cdots\\
    \cdots & H^{(1)}_{\bm k} & H^{(0)}_{\bm k} & H^{(-1)}_{\bm k} & \cdots \\
    \cdots & H^{(2)}_{\bm k} & H^{(1)} _{\bm k}& H^{(0)}_{\bm k} + \hbar\omega  & \cdots \\
    \ \rotatebox{315}{{\vdots}} & \vdots & \vdots & \vdots & \ddots
    \end{pmatrix}.\label{fhk}
\end{equation}
As a result, the bulk quasi-energy bands and eigenstates can be calculated by
\begin{equation}
    \mathcal{H}_{\bm k}\vert\boldsymbol{\varphi}_{\beta}\rangle = \beta(\bm{k}) \vert\boldsymbol{\varphi}_{\beta}(\bm{k})\rangle.
\end{equation}
 For each $\bm{k}$, the two centralmost quasi-energies and their associated eigenstates are adopted for subsequent analysis, see Fig.~1(c) of the main text and Figs.~\ref{fig1}(c, d).
  
\subsection{Berry flux}

The Berry flux $F(\bm{k})$ at a given momentum $\bm{k}$ is computed via Wilson loop formula,
\begin{align}
    F(\bm{k}) = -\arg\Big[ & 
        \langle \bm{\varphi}_{\beta} (\bm{k}) | \bm{\varphi}_{\beta} (\bm{k}+d\bm{k}_1) \rangle
        \langle \bm{\varphi}_{\beta} (\bm{k}+d\bm{k}_1) | \bm{\varphi}_{\beta} (\bm{k}+d\bm{k}_1+d\bm{k}_2) \rangle\nonumber\\
        &\times\langle \bm{\varphi}_{\beta} (\bm{k}+d\bm{k}_1+d\bm{k}_2) | \bm{\varphi}_{\beta} (\bm{k}+d\bm{k}_2) \rangle
        \langle \bm{\varphi}_{\beta} (\bm{k}+d\bm{k}_2) | \bm{\varphi}_{\beta} (\bm{k}) \rangle
    \Big],
    \label{BF}
\end{align}
where $d\bm{k}_1$ and $d\bm{k}_2$ represent infinitesimal displacement vectors used for uniform sampling in the Brillouin zone. Examples of $F(\bm k)$ are shown in Fig.~1(d) of the main text and the insets in Figs.~\ref{fig1}(c, d).

\subsection{Chern number}

We compute the Chern number by integrating the Berry flux over the entire BZ,
\begin{equation}
    C = \frac{1}{2\pi}\sum_{\bm{k}\in{\rm BZ}} F(\bm{k}),
    \label{Chern}
\end{equation}
where the sum represents a numerical integration over the discretized $\bm{k}$-points in the BZ. This way is used throughout the main text, for example in Figs.~2(d, e).

Alternatively, in Floquet systems, the effective Floquet Hamiltonian \(H_{\rm eff}\) provides an alternative approach to computing quasi-energies and eigenstates. It is defined as
\begin{equation}
    H_{\text{eff}} = \frac{i}{Z} \log[U(Z)],\quad
    H_{\text{eff}}\vert\Psi_{\beta}\rangle = \beta|\Psi_{\beta}\rangle,
    \label{Heffect}
\end{equation}
where \(U(Z)\) denotes the Floquet evolution operator over one full period \(Z = 2\pi/\Omega\). This operator is approximated by discretizing the period into \(N\) small intervals of length \(\Delta z = Z/N\): 
\begin{equation}
    U(Z) = \prod_{n=1}^{N}e^{-i H(Z-n\Delta z)\Delta z},
    \label{U_effect}
\end{equation}
with \(N = 100\) adopted in this work.

For large-scale systems, obtaining a Fourier expansion with sufficiently high precision is often limited in practice by computational resource constraints. When the primary focus is on capturing long-term stroboscopic dynamics over many periods, the effective Hamiltonian provides a powerful framework for computing the dynamical evolution from $z_0$ to $z_0 + nZ$ \cite{Bukov_2015_Universal}, that is,
\begin{equation}
    |\Psi(z_0+nZ)\rangle = e^{-i n H_{\text{eff}}  Z} |\Psi(z_0)\rangle
    \label{Heff evolution}
\end{equation}
This is the method used for the finite system in Figs.~4(a-c). 

To cross-validate the Chern number results, we perform the calculation using the effective Hamiltonian approach. We first construct the time-evolution operator $U(\bm{k},Z)$ from $H_{\bm{k}}(z)$, and then derive the effective Hamiltonian along with its eigenstates $\vert\boldsymbol{\Psi}_{\beta}(\bm{k})\rangle$ via Eq.~\eqref{Heffect}. Finally, we substitute these eigenstates for $\vert\boldsymbol{\varphi}_{\beta}(\bm{k})\rangle$ in Eq.~\eqref{BF} to compute the Chern number $C$ by Eq.\eqref{Chern} and the valley Chern number $C_V$ by Eq.~(3) of the main text. The corresponding results are presented in Figs.~\ref{fig2}(a, b), which are fully consistent with those in Figs.~2(d, e) of the main text.

Real space Chern number is calculated by Kitaev formula \cite{Rudner_2013_PRX_Anomalous,He_2019_PRB_Quasicrystalline,Chen_2023_PRB_Kitaev}:
\begin{equation}
C_K = 12\pi i \sum_{i \in A} \sum_{j \in B} \sum_{k \in C} \left( P_{ij} P_{jk} P_{ki} - P_{ik} P_{kj} P_{ji} \right),
\label{eq:kitaev_realspace_chern}
\end{equation}
Three subregions labeled by $A,B,C$ with sites in them by $i,j,k$ are illustrated in Fig.~\ref{Chern_real}(a), where $P_{ij} = \sum_{\beta < E_F} \Psi_\beta(\mathbf{r}_i) \Psi_\beta^*(\mathbf{r}_j)$ if $\mathbf{r}_i \in A $ and $ \mathbf{r}_j \in B$. $E_F$ is the Fermi energy and $\Psi_\beta (r) $ represents the value of eigenstate of the quasi-energy $\beta$ at position $r$ in real space. This Kitaev formula confirms that the Chern number is zero for the finite lattices used in Figs.~3 and 4 of the main text.

\subsection{Winding number}

In Floquet systems, another important topological invariant is the winding number, which can be used to identify anomalous Floquet edge states and is defined as \cite{Rudner_2013_PRX_Anomalous}:
\begin{equation}
    W_\beta = \frac{1}{8\pi^2}\int_0^Z dz \int_{\mathrm{BZ}} d\bm{k} \, \mathrm{Tr}\left[ U_{\beta}^{-1} \partial_z U_{\beta} \cdot (U_{\beta}^{-1} \partial_{k_x} U_{\beta}, U_{\beta}^{-1} \partial_{k_y} U_{\beta}) \right],
\end{equation}
where
\begin{align}
U_\beta(\bm{k}, z) =& 
\left\{
\begin{array}{ll}
U(\bm{k}, 2z), & \mbox{if } 0 \leq z \leq Z/2, \\
V_\beta(\bm{k}, 2Z - 2z), & \mbox{if } Z/2 \leq z \leq Z,
\end{array}
\right.
\label{winding}\\
U(\bm{k},z) =&\ \prod_{n=1}^{z/\Delta z}e^{-i H_{\bm {k}}(z-n\Delta z)\Delta z},\\ V_\beta(\bm{k}, z) =&\ e^{-i H_{\rm eff}(\bm{k}) z}, \\ 
\qquad
H_{\rm eff}(\bm{k}) = &\ \frac{i}{Z} \log U(\bm{k}, Z).
\end{align}
In calculation, the subscript \(\beta\) in \(V_{\beta}\) specifies the branch cut of the logarithm:
\begin{equation}
\begin{array}{l}
\log e^{-i\beta Z + i0^-} = -i \beta Z, \\[6pt]
\log e^{-i\beta Z + i0^+} = -i \beta Z - 2\pi i.
\end{array}
\label{winding_branch}
\end{equation}
Both \(W_\pi\) and \(W_0\) are analyzed. The \(W_\pi\) winding number vanishes throughout the entire parameter space. The distribution of \(W_0\) in the \(R\)–\(\delta\) parameter plane is illustrated in Fig.~\ref{fig2}(c), which exhibits a consistent profile with the Chern number distribution. Both the Chern and winding numbers vanish for the proposed edge states in the continuum (EICs). These results strictly exclude the existence of anomalous Floquet topological edge states within our studied system.

\newpage

\section{List of supplemental figures}

%% fig1
\begin{figure}[h]
\includegraphics[width=0.6\textwidth]{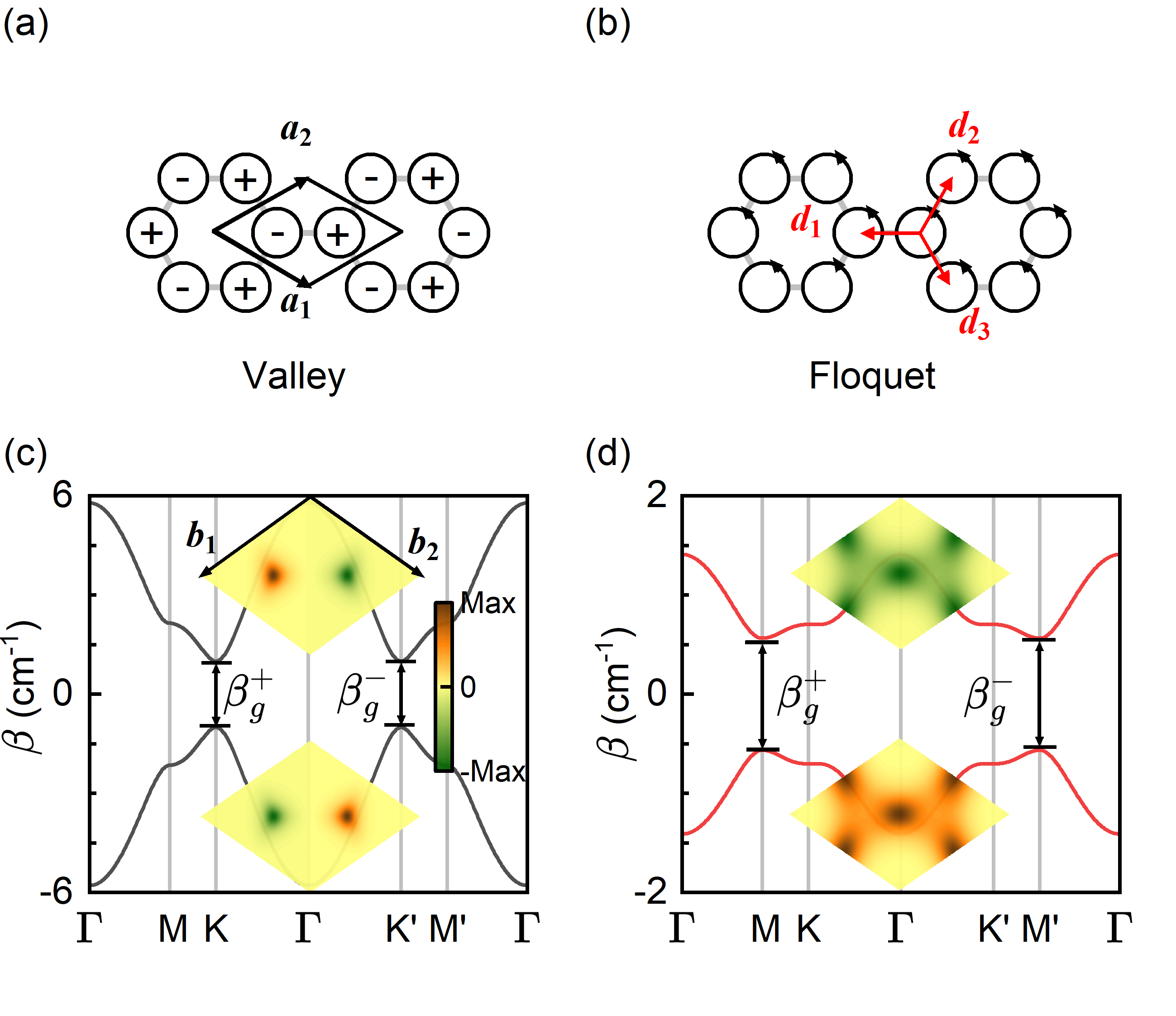}
\caption{
(a) Valley degree of freedom originating from sublattice potential modulation in a honeycomb lattice. $\bm{a}_1$ and $\bm{a}_2$ are the primitive lattice vectors with lattice constant $a$, and the $\pm$ symbols represent the site potential of $\pm\delta$. 
(b) Floquet degree of freedom implemented via helical waveguides in a honeycomb lattice. The circular arrows denote the out-of-plane helical rotation orientation of individual waveguides. $\bm{d}_1$, $\bm{d}_2$, and $\bm{d}_3$ represent the nearest-neighbor bonding vectors with a uniform length of $\frac{\sqrt{3}}{3}a$. 
(c) Quasi-energy band structure of the valley photonic crystal corresponding to the lattice in (a), calculated at $\delta=1\ \mathrm{cm}^{-1}$ and $R=0$. 
(d) Quasi-energy band structure of the Floquet photonic crystal corresponding to the helical lattice in (b), calculated at $R=6\ \mu\mathrm{m}$ and $\delta=0$. 
The insets in (c) and (d) plot the Berry curvature distributions of the respective band structures, where $\bm{b}_1$ and $\bm{b}_2$ denote the reciprocal lattice vectors.}
\label{fig1}
\end{figure}

The valley PhC in Fig.~\ref{fig1}(c) preserves time-reversal symmetry but breaks spatial inversion symmetry. In contrast, the Floquet PhC in Fig.~\ref{fig1}(d) breaks time-reversal symmetry while retaining spatial inversion symmetry. Despite the distinct symmetry breaking mechanisms, both systems exhibit even-parity band structures that satisfy $\beta(-\bm k)=\beta(\bm k)$. Nevertheless, their Berry flux distributions follow fundamentally different parity behaviors. The valley PhC supports odd-parity Berry flux obeying $F(-\bm k)=-F(\bm k)$, whereas the Floquet PhC features even-parity Berry flux with $F(-\bm k)=F(\bm k)$. These symmetry-dependent characteristics are fully consistent with the discussions in the main text.

% fig2
\newpage
\begin{figure}[h]
    \includegraphics[width=0.9\textwidth]{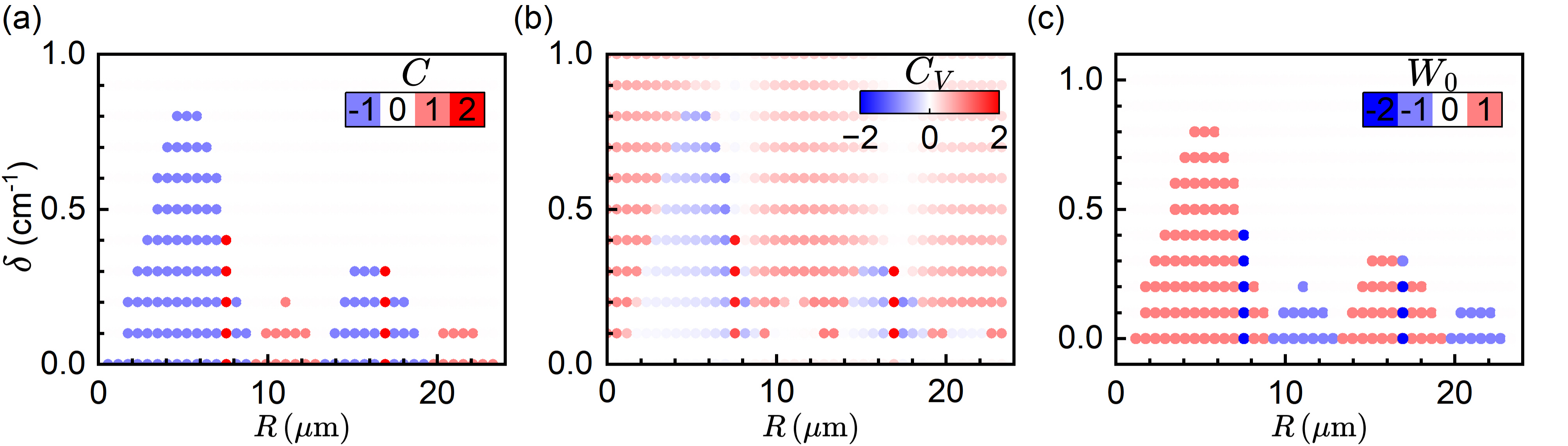}
    \caption{Topology invariants calculated with effective Hamiltonian. (a) Chern number \(C\), (b) Valley Chern number \(C_V\), and (c) winding number \(W_0\) in the \(R\)-\(\delta\) parameter space.}\label{fig2} 
\end{figure}

The Chern and valley Chern numbers presented in Figs.~2(d,e) of the main text are calculated directly from the original lattice Hamiltonian \({\cal H}_{\bm k}\) in Eq.~\eqref{fhk}. In comparison, the invariants in Figs.~\ref{fig2}(a,b) are evaluated using the effective Hamiltonian in Eq.~\eqref{Heffect}. The two independent calculation strategies yield exactly identical topological invariants, with the distribution profiles in Fig.~\ref{fig2}(a,b) perfectly consistent with those in Figs.~2(d,e). Moreover, the distribution of the winding number \(W_0\) obtained from the effective Hamiltonian also exhibits consistent variation with the Chern and valley Chern numbers, as verified by the comparison between Fig.~\ref{fig2}(c) and Figs.~\ref{fig2}(a, b). Such cross-validation of multiple topological invariants confirms that the proposed EICs originate from a non-topological mechanism.

% fig3
\newpage
\begin{figure}[h]
\includegraphics[width=0.75\textwidth]{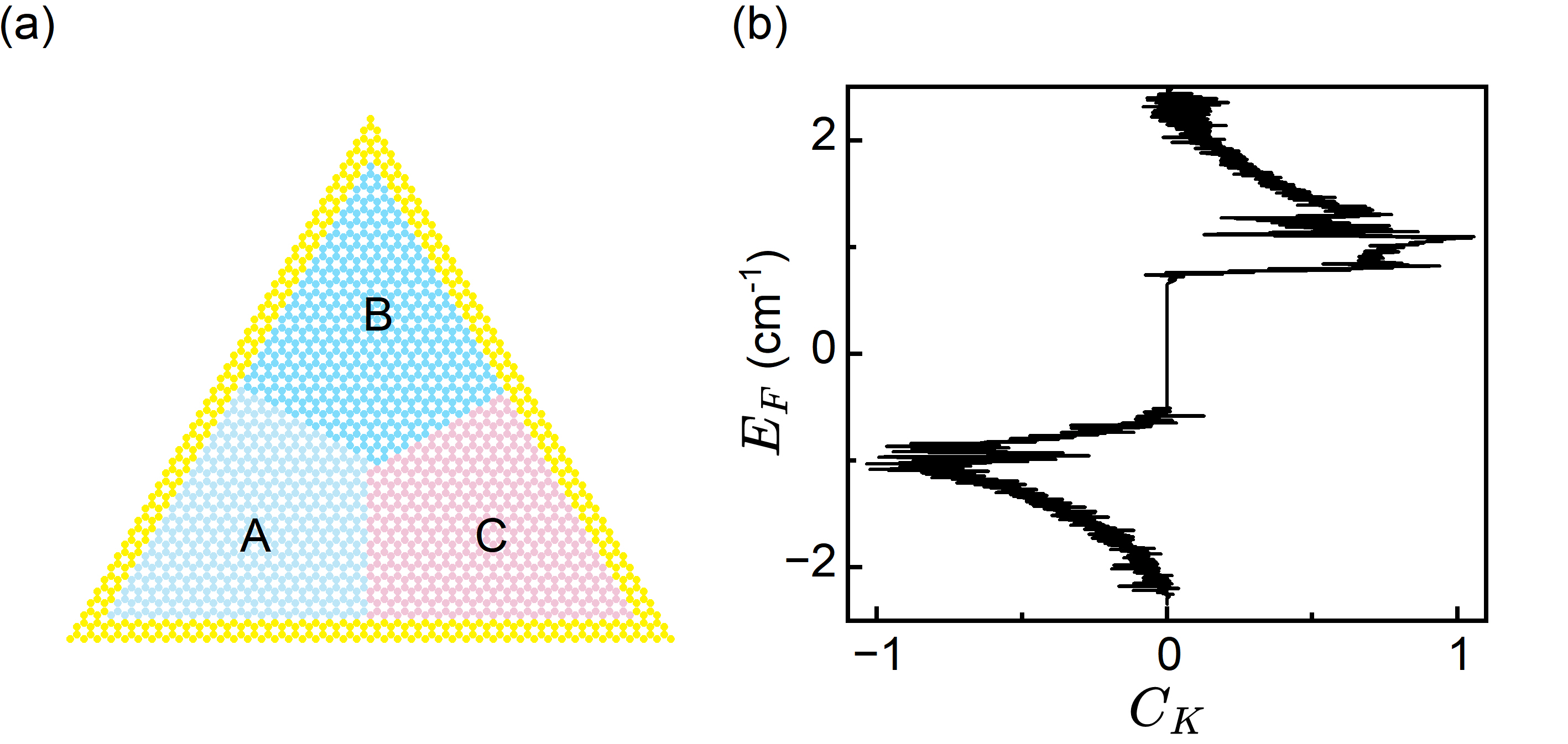}
 \caption{Chern number calculation for the finite lattice system. (a) Schematic of the Floquet lattice, which is identical to the structure in Fig.~3(c) of the main text. The entire lattice contains 2025 sites and is divided into three subregions for Chern number computation via the Kitaev formula, where the outermost four layers of edge sites are excluded to eliminate boundary effects. (b) Calculated Chern number as a function of the Fermi level \(E_F\). The Floquet lattice parameters are consistent with those adopted in Fig.~3(c) of the main text.}\label{Chern_real}
\end{figure}
 
To verify the trivial topological nature of the finite lattice utilized in Figs.~3 and 4 of the main text, we numerically calculate the Chern number using the Kitaev formula. The calculated results are plotted in Fig.~\ref{Chern_real}(b), which clearly demonstrates that the Chern number is zero for the studied lattice. This further validates the non-topological origin of the observed robust EIC transport.
 
% fig4
\newpage
\begin{figure}[h]
\includegraphics[width=0.75\textwidth]{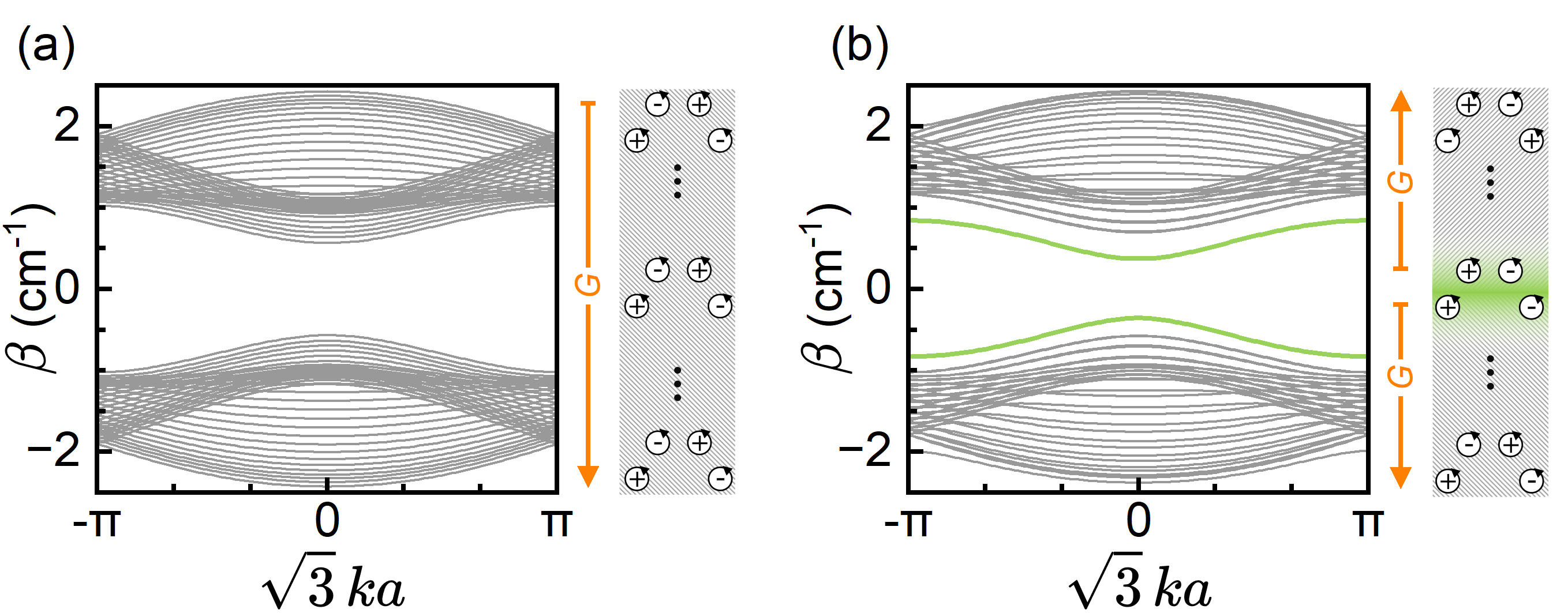}
\caption{(a) Band dispersion for an armchair edge without valley interface. (b) Band dispersion for an armchair edge integrated with a valley interface highlighted by the green shaded region. The parameters used here are $R=11.64\ \mu\mathrm{m}$ and $\delta=0.8\ \mathrm{cm}^{-1}$.}\label{fig4}
\end{figure}

As discussed in the main text, the proposed EICs can propagate along zigzag interfaces and cannot be sustained on armchair edges, as validated by the band structures in Fig.~\ref{fig4}. In sharp contrast, valley edge states are well supported on armchair interfaces in the presence of a valley domain wall, as illustrated in Fig.~\ref{fig4}(b). Such disparate edge-supporting behaviors distinctly differentiate the non-topological EICs from conventional topological valley edge states, confirming the corresponding discussions in the main text.

% fig5
\newpage
\begin{figure}[h]
\includegraphics[width=0.7\textwidth]{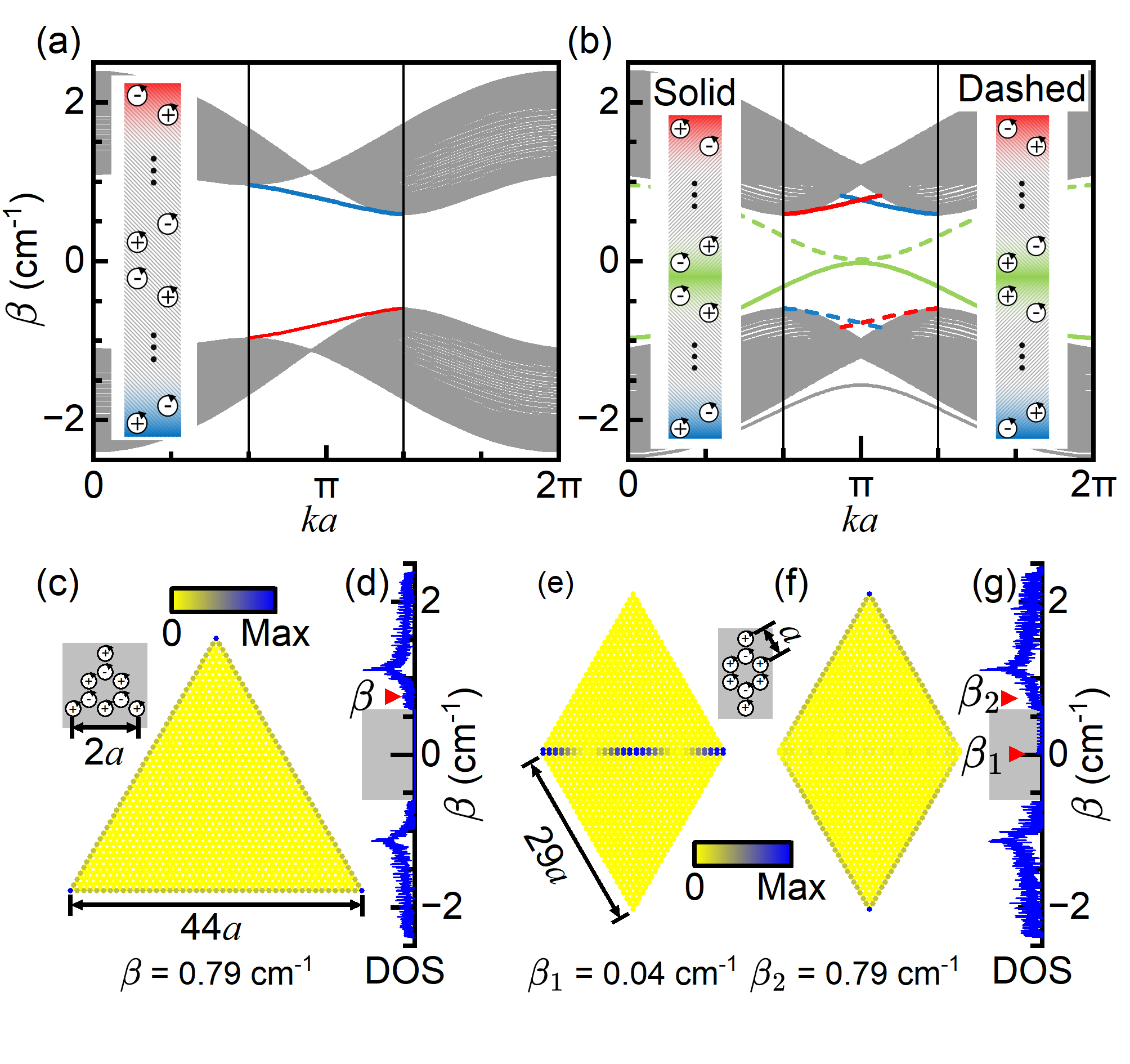}
\caption{Dispersion characteristics and spatial distributions of edge states in Floquet PhCs with $G=0$. (a, b) Band dispersions for Floquet PhCs without and with a valley interface, respectively, with corresponding structural insets. In (b), solid and dashed curves correspond to the left and right structural insets, respectively. The colored regions in the insets of (a) and (b) mark the spatial locations of the corresponding EIC modes. (c) Spatial distribution of the EIC at $\beta=0.79\,\text{cm}^{-1}$ in a equilateral triangular PhC and (d) the corresponding density of states (DOS). (e, f) Spatial distributions of two modes at $\beta_1=0.04\,\text{cm}^{-1}$ and $\beta_2=0.79\,\text{cm}^{-1}$ in a rhombic PhC, and (g) the corresponding DOS profile. The gray shaded triangular and rhombic insets illustrate the lattice geometries adopted in (c-g). The lattice contains 64 sites in (a, b), 2025 sites in (c, d), and 1800 sites in (e-g). All panels adopt identical structural parameters: $R=11.64\,\mu\text{m}$ and $\delta=0.8\,\text{cm}^{-1}$.}\label{fig5}
\end{figure}

We repeat the numerical calculations of Fig.~3 of the main text under identical parameters while setting $G=0$. The corresponding results are presented in Fig.~\ref{fig5}. The overall band dispersion features and spatial mode distributions show excellent consistency with those in Fig.~3 of the main text, which also confirms our related discussions in the main text.

% fig6
\newpage
\begin{figure}[h]
\includegraphics[width=0.99\textwidth]{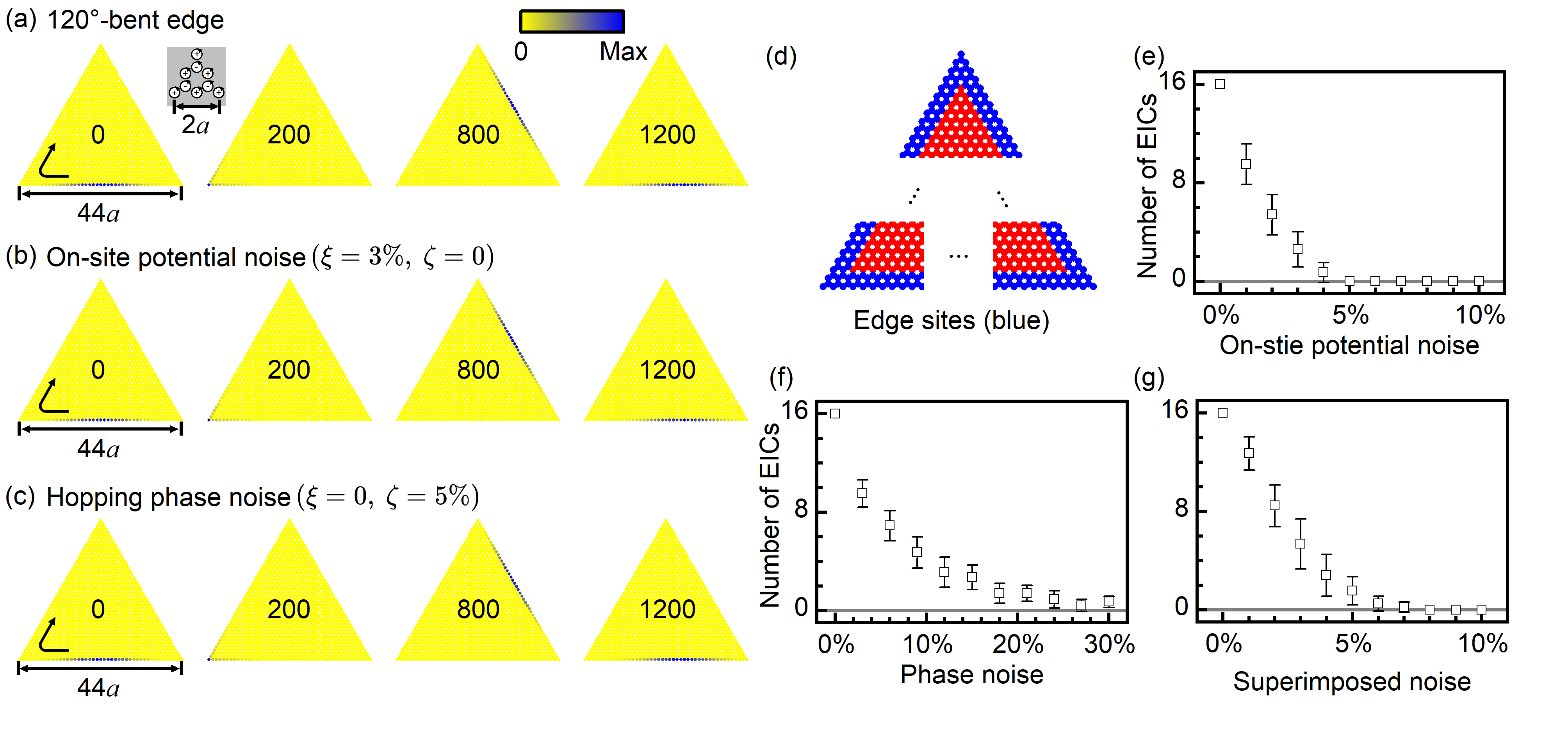}
\caption{Propagation of Gaussian-enveloped EICs in equilateral triangular Floquet PhCs with $G=0$, under different perturbation conditions. (a) Bent-corner propagation without disorder ($\xi=\zeta=0$). (b) Bent-corner with on-site potential noise ($\xi=3\%$, $\zeta=0$). (c) Bent-corner with hopping phase noise ($\xi=0$, $\zeta=5\%$). The triangular lattices adopted in (a)–(c) are identical to that in Fig.~\ref{fig5}(c). The EIC at $\beta=0.79\,\text{cm}^{-1}$ is modulated into a Gaussian pulse for all simulations. The numbers 0, 200, 840, and 1280 marked at the lattice centers denote the propagation lengths along the $z$-direction in units of $2\pi\Omega^{-1}$. (d) Schematic of edge sites defined as four layers of blue-colored sites. (e-g) Statistical numbers of sustained EICs as functions of individual on-site potential noise, individual hopping phase noise, and simultaneous dual disorder, respectively.} \label{fig6}
\end{figure}

Fig.~\ref{fig6} reproduces the simulations from Fig.~4 of the main text using identical parameters, except for \(G=0\). Each panel in Fig.~\ref{fig6} corresponds to its counterpart in Fig.~4 of the main text. The observed robustness of the EICs under this condition further supports the arguments presented in the main text.

% fig7
\newpage
\begin{figure}[h]
\includegraphics[width=0.9\textwidth]{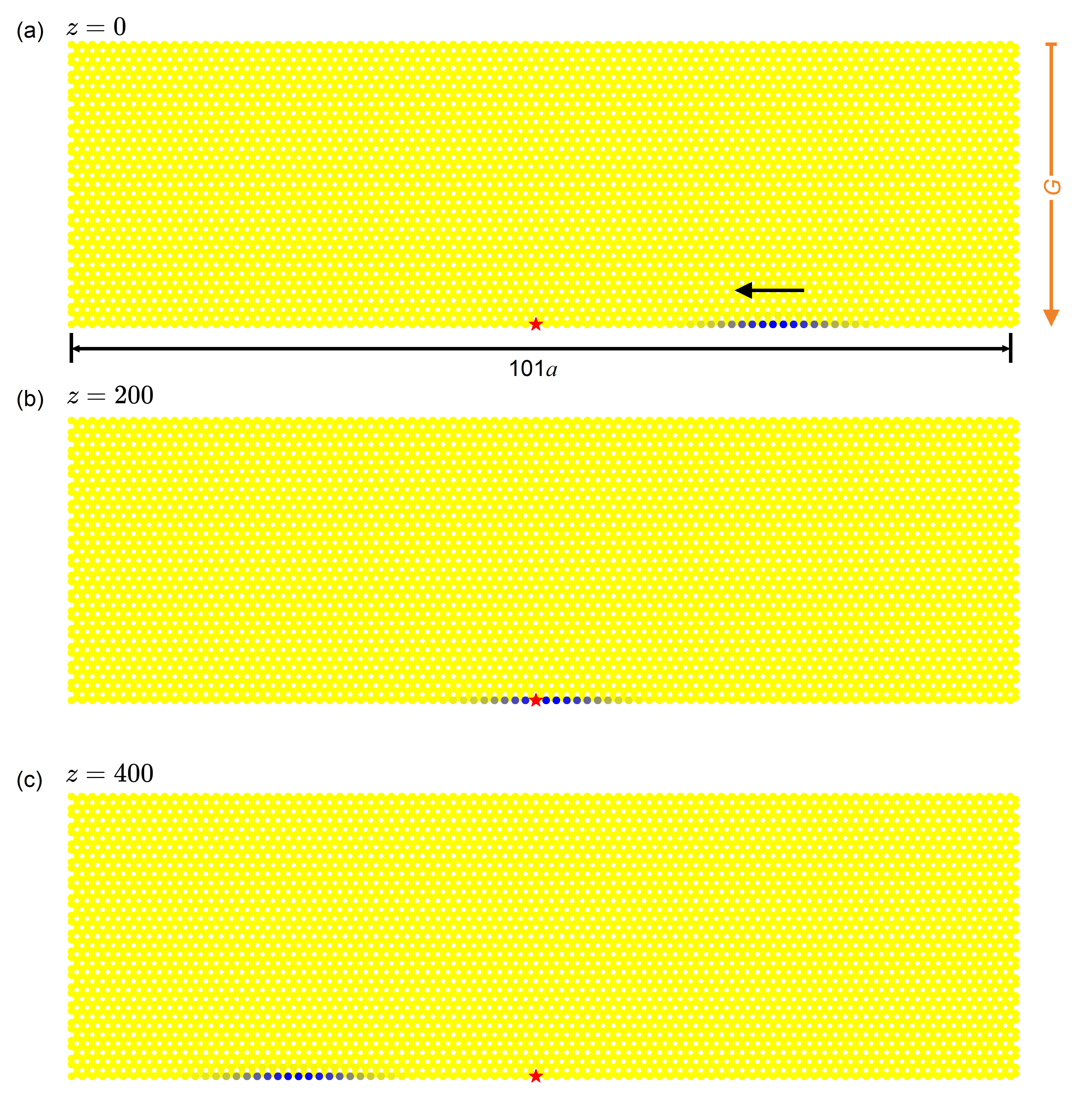}
\caption{Influence of a single local on-site potential defect on the propagation of the EIC with quasi-energy \(\beta=0.95\ \mathrm{cm}^{-1}\). The on-site potential at the lattice site marked by the red star is perturbed from its original value of \(0.9786\ \mathrm{cm}^{-1}\) to \(1.00796\ \mathrm{cm}^{-1}\), corresponding to a \(3\%\) potential deviation. Panels (a), (b), and (c) illustrate the field distributions at propagation lengths of \(z=0\), 200, and 400, in units of \(2\pi\Omega^{-1}\). All structural parameters are fixed to \(R=11.64\,\mu\text{m}\), \(\delta=0.8\,\mathrm{cm}^{-1}\), and \(G=4\,\mathrm{cm}^{-2}\).}\label{figs7}
\end{figure}

Fig.~\ref{figs7} illustrates the effect of a single local on‑site potential defect on the propagation of the EIC with quasi‑energy \(\beta=0.95\ \mathrm{cm}^{-1}\). The waveform remains nearly unchanged upon passing the defect site, confirming the robustness of the proposed EIC.

% fig8
\newpage
\begin{figure}[h]
\includegraphics[width=0.9\textwidth]{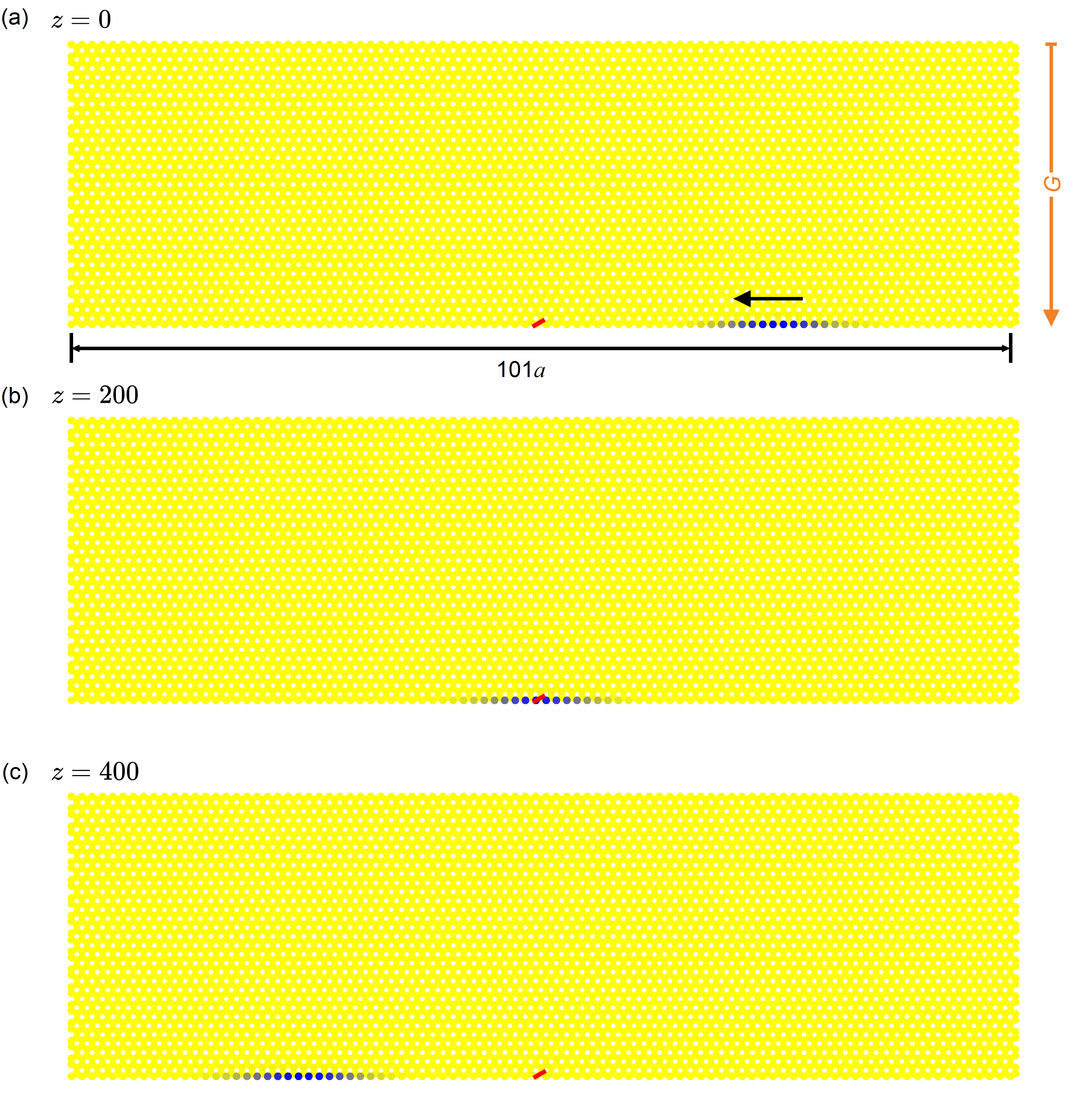}
\caption{Influence of a single local hopping phase defect on the propagation of the EIC with quasi-energy $\beta=0.95\ \mathrm{cm}^{-1}$. The hopping phase of the bond marked by the red rod is perturbed from the original value of $3.8\ \mathrm{rad}$ to $4.56\ \mathrm{rad}$, corresponding to a $20\%$ phase deviation. Panels (a), (b), and (c) illustrate the field distributions at propagation lengths of $z=0$, $200$, and $400$, in units of $2\pi\Omega^{-1}$. All structural parameters are fixed to $R=11.64\,\mu\text{m}$, $\delta=0.8\,\text{cm}^{-1}$, and $G=4\,\mathrm{cm}^{-2}$.} \label{figs8}
\end{figure}

Fig.~\ref{figs8} illustrates the effect of a single local hopping phase defect on the propagation of the EIC with quasi-energy \(\beta=0.95\ \mathrm{cm}^{-1}\). The waveform remains well preserved when passing the defective bond, further verifying the intrinsic robustness of the proposed Floquet-enabled EIC mode.

\newpage
\bibliography{SIref}% Produces the bibliography via BibTeX.